\documentclass[superscriptaddress,showkeys]{revtex4-2}
\usepackage{graphicx}% Include figure files
\usepackage{multirow}
\usepackage{lipsum}
\usepackage{placeins}% To stop the float
\usepackage{dcolumn}% Align table columns on decimal point
\usepackage{bm}% bold math
\usepackage[colorlinks=true, citecolor=red, urlcolor = violet, linkcolor= purple, bookmarks=true]{hyperref}
\usepackage{float}
\usepackage[utf8]{inputenc}
\usepackage{amsmath,amsfonts,amssymb}
\usepackage{natbib}

\usepackage{breakurl}
\usepackage{xcolor,verbatim,multirow}
\usepackage[normalem]{ulem}
\usepackage{subcaption} 
\usepackage{wrapfig}
\usepackage{adjustbox}
\usepackage{caption}
\usepackage{mathtools}
\usepackage{pgfplots}
\usepackage{mathrsfs}
\usepackage{booktabs}
\begin{document} 

\title[]{Parameter Estimation of Kerr-Bertotti-Robinson Black Holes Using Their Shadows}
\author{Heena Ali } \email{Heenasalroo101@gmail.com}
\affiliation{Centre for Theoretical Physics, Jamia Millia Islamia, New Delhi-110025, India }

\author{Sushant G. Ghosh}\email{sghosh2@jmi.ac.in}
\affiliation{Centre for Theoretical Physics, Jamia Millia Islamia, New Delhi-110025, India }
\affiliation{Astrophysics and Cosmology Research Unit, School of Mathematics, Statistics and Computer Science, University of KwaZulu-Natal, Private Bag X54001, Durban 4000, South Africa}
\begin{abstract}

 We investigate the shadow of Kerr-Bertotti-Robinson black holes (KBRBHs), which have a deviation parameter $B$ that captures the effect of an external magnetic field on the spacetime geometry. These spacetimes of Petrov type $D$ are asymptotically non-flat. We utilise the separability of the Hamilton-Jacobi equation to generate null geodesics and examine the crucial impact parameters for unstable photon orbits that define the black hole shadow.  We carefully investigate how the magnetic field strength $B$ and spin parameter $a$ influence black hole shadows, discovering that increasing $B$ increases the shadow size while also introducing additional distortions, especially at high spins.  We calculate the shadow observables, viz., area $A$ and oblateness $D$ and create contour plots in the parameter space $(a, B)$ to facilitate parameter estimation. We also investigate the dependence of the shadow on the observer's position, specifically by altering the radial coordinate $r_O$ and the inclination angle $\theta$.  For far viewers, the shadow approaches its asymptotic shape, but finite-distance observers perceive substantial deviations. The energy emission rate analysis reveals that the magnetic field parameter $B$ modifies the Hawking radiation spectrum, with increasing $B$ suppressing emission via backreaction, which lowers the Hawking temperature. Our findings confirm that KBRBH shadows encode imprints of magnetic deviations, thereby offering a potential avenue to distinguish Kerr from non-Kerr spacetimes and to probe magnetic effects in the strong-gravity regime.

\end{abstract} 

\maketitle

\section{Introduction}\label{Intro}

 The black hole shadows can be traced back to the early 70's with theoretical studies of photon orbits in a strong gravitational field, with seminal work by Synge \cite{Synge:1966okc} and Luminet \cite{Luminet:1979nyg}, who calculated the apparent shape of a Schwarzschild black hole and found circular photon orbits. Later, Bardeen \cite{Bardeen:1973tla} extended these results to Kerr black holes, revealing an abnormal, distorted shadow pattern due to frame-dragging effects. These important discoveries laid the theoretical groundwork for understanding how black holes' powerful gravitational fields deflect light, resulting in shadows. Modern astrophysics has enabled the study of black hole shadows, allowing for the testing of alternative gravity theories and general relativity in the strong field regime. Early theoretical work on shadows was investigated in rotating black holes \cite{Amarilla:2010zq,Amarilla:2011fx,Amarilla:2013sj}, and later extended to brane-world scenarios \cite{Amarilla:2011fx,Atamurotov:2013dpa} and Kaluza-Klein black holes \cite{Amarilla:2013sj}. Shadow analysis has been extended to Ho\v{r}ava--Lifshitz black holes \cite{Atamurotov:2013dpa}, traversable wormholes \cite{Nedkova:2013msa}, and non-Kerr spacetimes \cite{Atamurotov:2013sca}.

Recent investigations into shadow have explored higher-dimensional black holes \cite{Papnoi:2014aaa}, Einstein-Born-Infeld solutions \cite{Atamurotov:2015xfa}, and black holes surrounded by quintessential energy \cite{Abdujabbarov:2015pqp}. Attention has been given to rotating regular black holes \cite{Abdujabbarov:2016hnw} and their horizon structure, with significant work on $f(R)$ gravity models \cite{Dastan:2016vhb} and non-commutative geometries \cite{Sharif:2016znp,Ahmed:2025boj}. The shadow properties of Konoplya-Zhidenko black holes \cite{Wang:2017hjl} and Kerr-Newman spacetimes \cite{Tsukamoto:2017fxq} have provided crucial benchmarks for testing deviations from standard Kerr geometry,  whereas the Konoplya–Zhidenko metric, introduced in Ref. \cite{Konoplya:2016pmh}

The influence of surrounding matter fields on shadows has been extensively investigated, including anisotropic fluid \cite{Singh:2017xle,Kumar:2017tdw}, perfect fluid dark matter \cite{Haroon:2018ryd,Hou:2018avu,Haroon:2019new}, and global monopole  \cite{Haroon:2019new}. Other theoretical frameworks, incorporating asymptotically safe gravity models \cite{Kumar:2019ohr} and Newman-Janis Algorithm (NJA) constructions \cite{Shaikh:2019fpu}, have enabled systematic studies of shadow deformation. Notable contributions include polytropic black holes \cite{Contreras:2019nih}, scale-dependent scenarios \cite{Contreras:2019cmf}, and squashed Kaluza-Klein solutions \cite{Long:2019nox}.

Falcke \textit{et al.} \cite{Falcke:1999pj} first demonstrated the possibility of capturing an image of a black hole's shadow, particularly from Sgr A* at the centre of the Milky Way. This set the foundation for the Event Horizon Telescope (EHT) collaboration, which later turned this idea into reality by capturing famous images of M87* and Sgr A* \cite{EventHorizonTelescope:2019dse,EventHorizonTelescope:2022wkp}. Since then, the black hole shadow has served as a powerful tool for probing gravity in the strong-field regime, thereby testing general relativity and modified theories of gravity.

With in-depth studies of regular solutions \cite{Kumar:2019pjp}, Kalb-Ramond gravity models \cite{Kumar:2020hgm}, and astrometric observables \cite{Chang:2020miq}, charged spinning black holes have received special attention. The five-dimensional extensions have further enhanced the theoretical landscape \cite{Ahmed:2020jic}. With applications to photon rings \cite{Kumar:2020ltt} and anisotropic matter effects \cite{Badia:2020pnh}, the 4D Einstein-Gauss-Bonnet gravity \cite{Kumar:2020owy,Chen:2020aix, Kumar:2020ltt} has proven to be an especially successful method. Recent observational constraints from EHT have motivated studies connecting shadows with fundamental black hole properties. The shadow-quasinormal mode resemblance \cite{Jusufi:2020dhz,Yang:2021zqy} has been established, while applications to Einstein-Euler-Heisenberg theory \cite{Lambiase:2024lvo} and nonlinear electrodynamics \cite{Raza:2023vkn} have furnished new observational signatures. The effect of plasma \cite{Chowdhuri:2020ipb} and the cosmological constant \cite{Eiroa:2017uuq} has been incorporated into shadow analyses, with precise attention to observational degeneracies \cite{Xu:2018mkl}. The symmergent gravity \cite{Pantig:2022qak} and 5D black strings \cite{Tang:2022bcm} have developed a variety of testable circumstances, while rotating regular AdS black holes \cite{Singh:2022dqs} have been used to investigate the relationship between shadows and black hole thermodynamics. Recent work has focused on rotating black holes in massive gravity \cite{Zubair:2023krl}, general rotating solutions \cite{Meng:2023uws}, wormholes in plasma environments \cite{Kumar:2023wfp} and quantum-corrected metrics \cite{Ali:2024ssf}. Einstein-Maxwell scalar theory \cite{Yunusov:2024xzu} has provided additional avenues for testing fundamental physics.
Black holes in the magnetic field are fundamental to relativistic astrophysics, connecting theoretical predictions with observational phenomena. The importance of understanding black holes in magnetic fields has been emphasised by the recent LIGO \cite{LIGOScientific:2016aoc} detections of gravitational waves from binary black hole mergers and the EHT's \cite{EventHorizonTelescope:2019dse, EventHorizonTelescope:2021bee, EventHorizonTelescope:2022wkp} seminal images of supermassive black hole shadows. The spacetime geometry and observational properties of astrophysical black holes are distinct due to the presence of magnetized plasma, whereas the Kerr metric defines vacuum black holes \cite{Griffiths:2009dfa,Kerr:1963ud,Stephani:2003tm}.

The accurate solution to the Einstein-Maxwell equations describing a rotating black hole confined in a uniform magnetic field, which we call  Kerr-Bertotti-Robinson black holes (KBRBH) \cite{Podolsky:2025tle}, is a breakthrough.  Unlike the Kerr-Melvin spacetime \cite{Ernst:1976mzr, Ernst:1976bsr}, which is of Petrov type I and features unbounded ergoregions, the KBRBH solution is of Petrov type D with bounded ergoregions and an asymptotically uniform magnetic field \cite{Podolsky:2025tle}. These properties make it a more natural model for astrophysical black holes, as test particles can escape to infinity, and spacetime lacks the pathological features of earlier solutions \cite{Podolsky:2025tle, Zeng:2025olq}.
 Recent studies have explored its thermodynamic properties \cite{Podolsky:2025tle}, energy extraction via magnetic reconnection \cite{Zeng:2025olq}, and geodesic structure \cite{Wang:2025vsx}. Notably, the KBRBH shows $a$ Meissner-like effect, where the magnetic field is expelled from the horizon for extremal configurations \cite{Podolsky:2025tle}, akin to the behaviour observed in Kerr-Melvin spacetimes \cite{Bicak:2015lxa, Gurlebeck:2018smy}. The shadow characteristics and photon orbits of a related magnetized solution were also recently investigated in the Kerr-Bertotti-Robinson spacetime \cite{Zeng:2025tji}.
This paper investigates the parameter estimation of KBRBHs using their shadows and constraints from EHT observations. The shadow—a dark region surrounding the black hole is shaped by photon orbits and serves as a probe of spacetime geometry \cite{Gralla:2019xty, Solanki:2022glc}. For the KBRBH, the shadow deviates from the Kerr case due to the magnetic field’s influence, with the deformation \cite{Wang:2025vsx}. We employ ray-tracing procedures to compute shadows for KBRBHs and perform a comparison with Kerr shadows.
Our analysis reveals that the magnetic field suppresses shadow asymmetry for moderate $B$ but amplifies deviations near extremality \cite{Wang:2025vsx}. 

The structure of this paper is as follows. We begin in Sec.~\ref{Sec2} with the KBRBH metric and discuss generic features of the black hole, including horizon structures and thermodynamical variables.  Sec.~\ref {Sec3} and Sec.~\ref {Sec4}  focus on the null geodesics and black hole shadows, respectively. We also discuss how the parameter $B$ affects their shape and size. In Sec .~\ref {Sec5}, we provide the shadow characterisation observables and utilise them to estimate the parameters related to KBRBHs. We compute the energy emission rate in Sec.~\ref{EE rate}. Finally, we present our conclusions and discuss the implications of our findings in Section~\ref{Sec6}.

\section{ KBRBHs Metric in Boyer--Lindquist Coordinates}\label{Sec2}
The KBRBHs metric \cite{Podolsky:2025tle} in Boyer--Lindquist coordinates $(t, r, \theta, \phi)$ is :
	\begin{eqnarray}
	ds^2 &=& \frac{1}{\Omega^2} \left[ 
		- \left( \frac{Q}{\Sigma} - \frac{a^2 \sin^4\theta P}{\Sigma} \right) dt^2
		+ \frac{\Sigma}{Q} dr^2
		+ \frac{\Sigma}{P} d\theta^2
		+ \left( \frac{(r^2 + a^2)^2 \sin^2\theta P}{\Sigma} - \frac{Q a^2 \sin^4\theta}{\Sigma} \right) d\phi^2 \right. \nonumber \\
&&		\left. - \frac{2 a \sin^2\theta}{\Sigma} \left[ Q - (r^2 + a^2)P \right] dt \, d\phi 
		\right]\label{MKBH}
\end{eqnarray}
with the functions defined as:
	\begin{align}
		\Sigma &= r^2 + a^2 \cos^2\theta,~~~~~~P = 1 + B^2 \left( \frac{m^2 I_2}{I_1^2} - a^2 \right) \cos^2\theta,    \nonumber\\
		Q &= (1 + B^2 r^2)\Delta,~~~~~~\Omega^2 = (1 + B^2 r^2) - B^2 \Delta \cos^2\theta, \\
		\Delta &= \left(1 - B^2 m^2 \frac{I_2}{{I_1}^2} \right) r^2 - 2m \frac{I_2}{I_1} r + a^2, \nonumber \\
		I_1 &= 1 - \frac{1}{2} B^2 a^2,~~~~~~I_2 = 1 - B^2 a^2 . 
	\end{align} 
Here $m$ is the black hole mass, $a$ is the spin - $J/m$,  $B$ is the magnetic field parameter.
The KBRBH metric admits three physical parameters: mass $m$, spin $a$, and magnetic field strength $B$. For $B=0$, it reduces to the Kerr solution \cite{Kerr:1963ud}; for $m=0$, it becomes the Bertotti-Robinson universe \cite{Bertotti:1959pf,Robinson:1959ev}, and for $a=0$, it describes a magnetized Schwarzschild black hole \cite{Podolsky:2025tle}. 
It is important to emphasise that the parameter $B$ does not merely describe a test magnetic field in a Kerr background. Rather, it now enters the metric functions via the Einstein--Maxwell equations, thereby modifying the spacetime geometry itself \cite{Podolsky:2025tle}. This distinguishes the KBRBH from Kerr-plus-plasma models, where the background geometry remains unaffected by magnetic fields.

\begin{figure}[hbt!]
\includegraphics[scale=0.8]{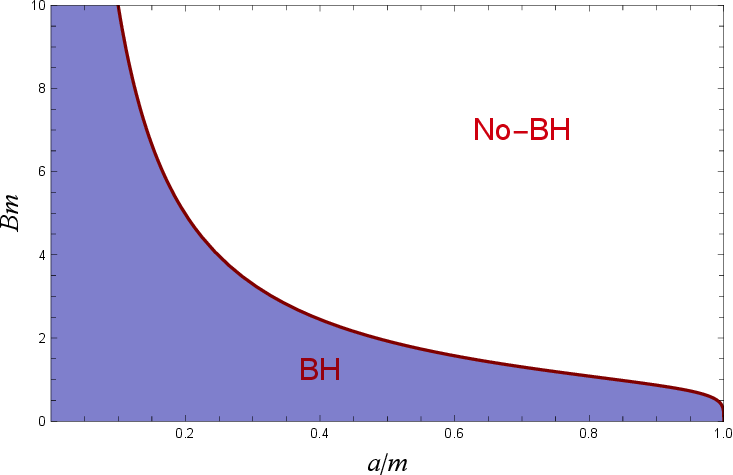}
\caption{\label{fig1} Parameter space in the $ (a, B) $ plane for 
KBRBHs. 
The shaded (blue) region corresponds to the existence of both inner and outer horizons.
The boundary curve (red) represents the extremal black hole limit, beyond which no real horizons exist, even though it comes to a naked singularity or no-horizon spacetime. 
Increasing magnetic field strength $B$ restricts the allowed values of the spin parameter $a$, thus tightening the conditions for black hole formation. }
\end{figure}

We depict the parameter space $(a, B)$ in Fig.~\ref{fig1}. The shaded region approximates a black hole with two horizons, while for all parameter values ($B_E$, $a_E$) along the dark red solid line, the radius $r_+=r_-$ degenerates, indicating an extremal black hole. In contrast, the white region represents spacetimes without horizons. 
 
Normalized by the black hole mass $m$, the magnetic field parameter $B$ denotes an external uniform field. For M87* ($m \approx 6.5 \times 10^9 M_\odot$), $Bm = 0.1$, in accordance with EHT polarization constraints \cite{EventHorizonTelescope:2021srq}. As it matches estimates from jet power \cite{Narayan:2003by} and remains below theoretical limitations for horizon stability (Fig.~\ref{fig1}), this range $Bm \in [0, 0.3]$ is astrophysically feasible.   Despite the fact that real fields in accretion systems are not uniform \cite{Tchekhovskoy:2011zx}, our uniform approximation captures important nonlinear electrodynamical effects that are crucial to the formation of shadows \cite{Ernst:1976mzr}. 

As shown by the recent high-resolution polarimetric imaging by the EHT, magnetic fields are dynamically significant  on the event-horizon scales of supermassive black holes (SMBHs). The polarized emission for M87* is compatible with a magnetically arrested disk (MAD) configuration, indicating ordered magnetic fields near the horizon and in the synchrotron-emitting region with strengths in the range $B \sim 1$--$30~{\rm G}$ \cite{EventHorizonTelescope:2021srq,EventHorizonTelescope:2022xqj}. Horizon-scale ordered fields of comparable magnitude are also indicated by similar results for Sgr~A*. Depending on assumptions regarding jet power and plasma composition, models of the jet-launching funnel within the Blandford--Znajek framework suggest significantly stronger fields beyond the radiative region, with estimates extending to $B \sim 10^2$--$10^3~{\rm G}$ \cite{Gralla:2019xty}. These considerations emphasize the fact that realistic SMBHs are naturally embedded in magnetized environments rather than being isolated vacuum Kerr solutions. Only emissivity, opacity, and Faraday effects are used to incorporate magnetic fields in standard Kerr ray-tracing; the spacetime geometry itself is left unaltered.   However, in MAD-like states, the magnetic energy density directly contributes to the stress energy tensor by being comparable to or exceeding the gas pressure. These back-reaction effects are self-consistently preserved by KBRBHs, which are precise solutions of the Einstein-Maxwell equations. The deviation parameter $B$ provides a straightforward method for estimating the effects of external fields on parameter estimation, shadow morphology modification, and photon region perturbation.

The spacetime~\eqref{MKBH} is consistent with the Einstein–Maxwell equations. The 1-form gauge potential is given by:  
\begin{equation}
\textbf{A}=A_{\mu} \, x^{\mu} = \dfrac{{\rm e}^{\mathrm{i} \gamma}}{2B} \left[ \partial_r\Omega \, \dfrac{a \, t - (r^2 + a^2) \, \phi}{r + \mathrm{i} a \cos\theta} + \dfrac{\mathrm{i} \, \partial_\theta\Omega}{\sin\theta} \, \dfrac{ t - a \sin^2\theta \, \phi}{r + \mathrm{i} a \cos\theta} + (\Omega - 1) \, \phi \right],  
\label{Amu}
\end{equation}  
which vanishes in the limit $B = 0$. Here, $\mathrm{i}$ is the imaginary unit. The metric (\ref{MKBH}) satisfies $\textbf{A}^{real}=2Re\textbf{A} $ \cite{Podolsky:2025tle}, with $\textbf{A}^{real}$  corresponding to the physically meaningful electromagnetic potential. $\gamma$ represents an arbitrary phase parameter governing the relative contributions of the electric and magnetic components. In particular, setting $\gamma = 0$ yields a purely magnetic configuration, whereas $\gamma = \pi / 2$ corresponds to a purely electric field. Crucially, the choice of $\gamma$ has no influence on the geometry of the spacetime.
Furthermore, the metric belongs to Petrov type D—distinct from the type I structure of the Kerr–Melvin solution—and features a source-free electromagnetic field that is neither null nor aligned with the principal null directions \cite{Podolsky:2025tle}.

\paragraph{Horizon Structure of the KBRBH spacetime}
The existence of horizons in the KBRBH spacetime is determined by the condition that the metric function $\Delta(r)$ possesses real roots \cite{Ernst:1976bsr}. The function $\Delta(r)$ appears in the metric as
	\begin{equation}\label{Delta}
		\Delta(r) = \left(1 - B^2 m^2 \frac{I_2}{{I_1}^2} \right) r^2 - 2m \frac{I_2}{I_1} r + a^2,
	\end{equation}
	where the auxiliary functions are defined as
	\begin{align}
		I_1 &= 1 - \frac{1}{2} B^2 a^2, \\
		I_2 &= 1 - B^2 a^2.
	\end{align}
	
The radii of the outer and inner horizons of the KBRBH black hole are given by:
\[
r_{\pm} = \frac{m I_2 \pm \sqrt{m^2 I_2 - a^2 I_1^2}}{I_1^2 - B^2 m^2 I_2},
\]
The existence of horizons requires 
	\begin{equation}
		a^2 \leq m^2 \left( \frac{I_2}{{I_1}^2} \right)
		\label{eq:horizon_condition}
        \end{equation}
with equality corresponding to the extremal black hole \cite{Gibbons:2013yq,Aliev:2005bi}.

The extremal KBRBH with degenerate horizons can be obtained, where
\begin{equation}
    m^2I_2-a^2I_1^2=0\label{extermal}\end{equation}which gives
    \begin{equation}
        B^2 =\frac{2}{a^4} \left( m - \sqrt{m^2 - a^2} \right) \sqrt{m^2 - a^2}
    \end{equation}

\begin{figure*}[hbt!]
\centering
\begin{tabular}{c c}
    \includegraphics[scale=0.58]{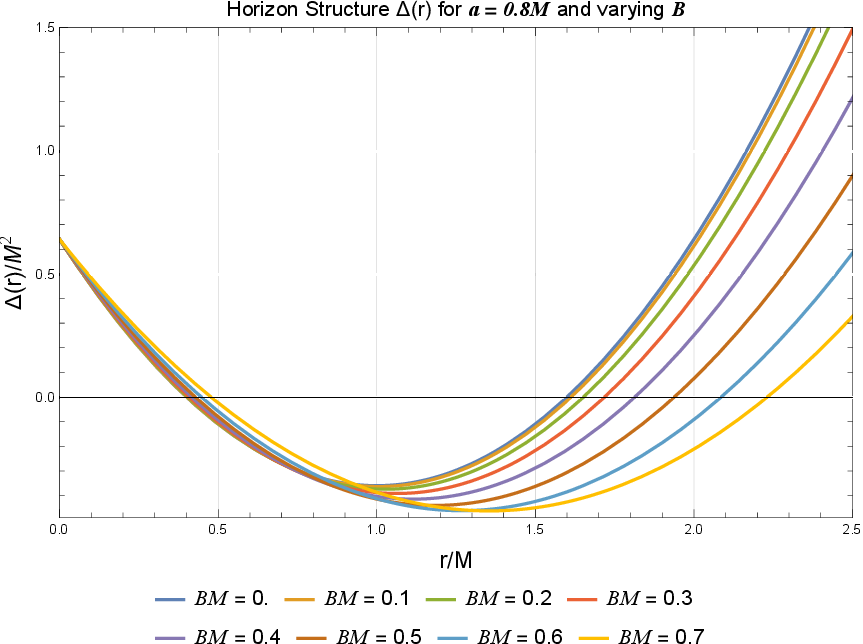}&
    \includegraphics[scale=0.58]{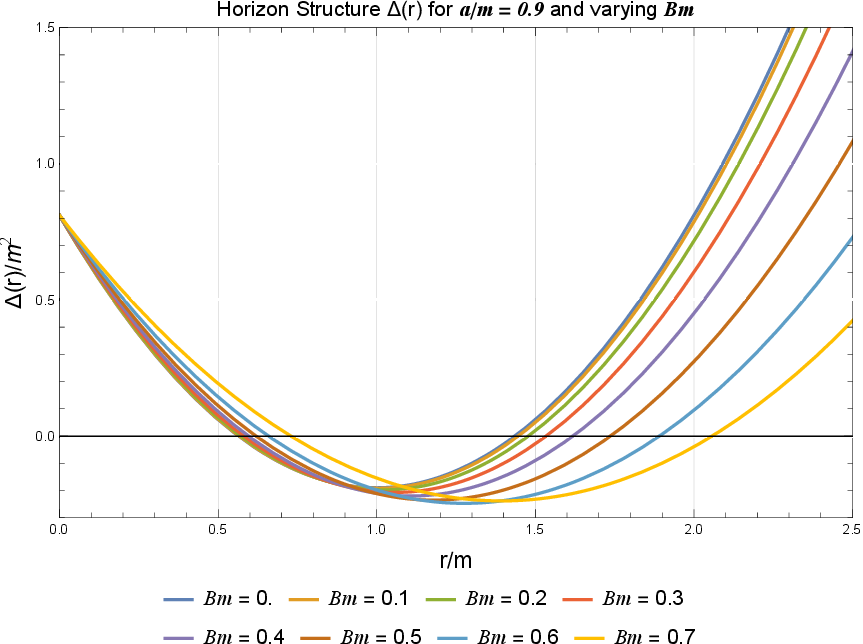}
    \end{tabular}
\caption{Comparison of the horizon structure between the KBRBHs and the Kerr black hole ($ B = 0 $) . 
The presence of a uniform magnetic field modifies the locations of the inner ($ r_- $) and outer ($ r_+ $) horizons. 
As  $B$ increases, the event horizons  ($ r_+$) shift outward.  }\label{horizons}

\end{figure*}

This equation specifies the allowable region in the $(a, B)$ parameter space where the KBRBH spacetime admits real horizons while avoiding naked singularities. As the magnetic field intensity $B$ increases, the maximum permitted value of the spin parameter $a$ drops, with the extremal black hole limit determined by the saturation of Eq.~\eqref{eq:horizon_condition}. The KBRBH horizon depicted in Fig.~\ref{horizons} demonstrates that for given values of $B$, there exists a critical extremal value $a_E$ of $a$, and similarly, a critical extremal value $B_E$ of $B$ for a given value of $a$, where $\Delta=0$ has a double root.

\paragraph{Stationary Limit Surface.}
The stationary limit surface (SLS), which marks the boundary of the ergoregion, 
is defined by the condition $g_{tt}=0$ \cite{1972ApJ...178..347B}. Explicitly,
\begin{equation}
g_{tt} = \frac{Q(r)}{\Sigma} - \frac{a^2 \sin^4\theta \, P(\theta)}{\Sigma} = 0.
\end{equation}
The solution of this equation yields the surface 
\begin{equation}
r = r_{\text{SLS}}(\theta),
\end{equation}
which lies outside the event horizon for all $\theta$ except at the poles 
($\theta=0,\pi$), where the two surfaces coincide. 
The region bounded by the SLS and the horizon is the ergoregion, 
where particles cannot remain static due to frame dragging 
\cite{Misner:1973prb,Chandrasekhar:1985kt}.

\paragraph{Surface Gravity and Hawking Temperature.}
The Hawking temperature has previously been discussed \cite{Podolsky:2025tle}. However, since Hawking temperature is required later in the energy emission rate calculation and to make present work self contained, we briefly discuss the relevant results here. The surface gravity $\kappa$ of the rotating KBRBH is obtained from 
the standard Killing definition \cite{Wald:1984rg}:
\begin{equation}
\kappa = \left. \frac{1}{2(r^2 + a^2)} \frac{dQ}{dr} \right|_{r = r_+},
\end{equation}
Applying ${{Q}(r_+)=0}$, we get
\begin{align}
\kappa&= \frac{1+B^2r_+^2}{{r_{+}^2+a^2}}\,
    \Big( m\,\frac{I_2}{I_1} -\frac{a^2}{r_+}\Big )\,. \label{kappa-h}
\end{align}

Accordingly, the Hawking temperature is
\begin{equation}
T_H = \frac{\kappa}{2\pi} 
    = \frac{1+B^2r_+^2}{2\pi(r_{+}^2+a^2)}\Big( m\,\frac{I_2}{I_1} -\frac{a^2}{r_+}\Big) \Big|_{r=r_+}.
\end{equation}
Thus, the presence of the magnetic deviation parameter $B$ modifies the surface gravity. This behaviour is consistent with earlier analyses of magnetised Kerr and Kerr--Newman--Melvin black holes
\cite{Gibbons:2013yq,1985JMP....26..155G}.

\paragraph{Frame-Dragging Angular Velocity.}
The angular velocity of ZAMOs, due to frame dragging, reads\cite{Kumar:2019pjp,Kumar:2020hgm}
\begin{equation}
\omega = -\frac{g_{t\phi}}{g_{\phi\phi}}
       = \frac{a \left[ Q(r) - (r^2 + a^2) P(\theta) \right]}
              {(r^2 + a^2)^2 P(\theta) - Q(r) a^2 \sin^2\theta},
\end{equation}
which measures the dragging of inertial frames caused by black hole spin \cite{1972ApJ...178..347B,Chandrasekhar:1985kt}, 
that yields the corresponding expression, in the $B \to 0$ limit, for the Kerr black hole  \cite{Kumar:2020hgm}.

\paragraph{Area Law and Bekenstein--Hawking Entropy.}
The entropy and area law were also previously discussed \cite{Podolsky:2025tle}. Here we include a concise summary to ensure completeness. Including the conformal factor $\Omega^-2$ in the KBRBH metric, the horizon area reads
\begin{equation}
{A}_H = \int_0^{2\pi} d\phi \int_0^{\pi} d\theta 
   \, \sqrt{g_{\theta\theta} g_{\phi\phi}} \Big|_{r=r_+}
     =\frac{1}{\Omega^2}\int_0^{2\pi C}\!\int_0^\pi \sqrt{\tilde{g_{\theta \theta}}\, \tilde{g_{\phi \phi}}}\,\,d \theta \, d \phi, 
\end{equation}
where $C$ is the conicity parameter \cite{Podolsky:2025tle} . The above integral, on using (\ref{MKBH}), simplifies to
\begin{equation}
{A}_H = 4\pi C\,\frac{r_+^2+a^2}{1+B^2r_+^2 },
\end{equation}
which reduces to the Kerr horizon area $A_{\rm Kerr} = 4 \pi C (r_+^2 + a^2)$ in the absence of the magnetic field. The horizon area and Bekenstein–Hawking entropy KBRBH are altered because of the uniform magnetic field $B$.

According to the Bekenstein--Hawking relation \cite{Bekenstein:1973ur,Hawking:1975vcx,Wald:1984rg}, the entropy is
\begin{equation}
S = \frac{{A}_H}{4} = \pi C\,\frac{r_+^2+a^2}{1+B^2r_+^2 }.
\end{equation}

The magnetic field effectively reduces the horizon area and entropy, reflecting the fact that the spacetime is “compressed” by the conformal factor. These corrections are small for weak fields ($B \ll 1$), while the decrease becomes significant for stronger fields.

\begin{figure*}[hbt!]
\centering
\begin{tabular}{c c c c}
    \includegraphics[scale=0.5]{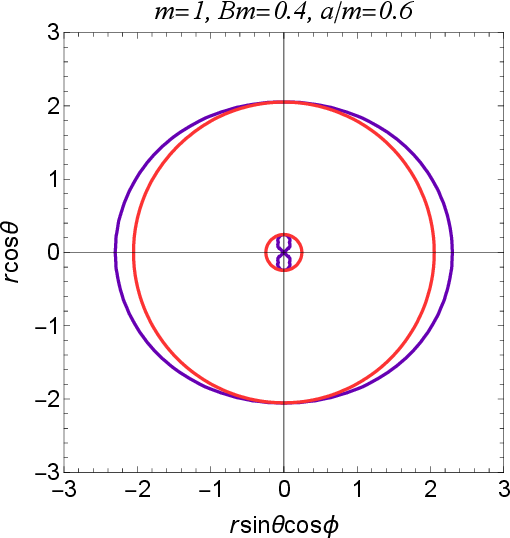}&
    \includegraphics[scale=0.5]{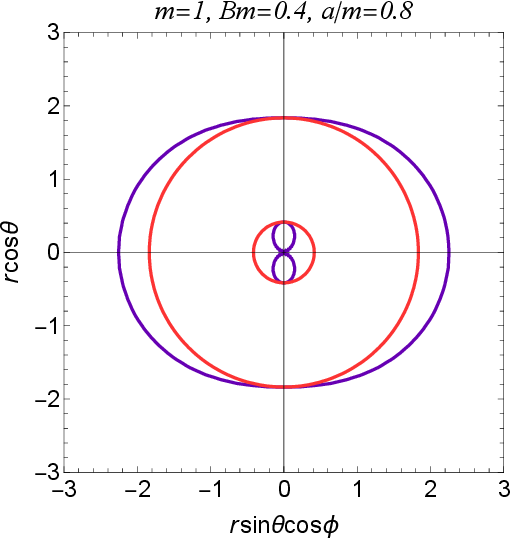}&
    \includegraphics[scale=0.5]{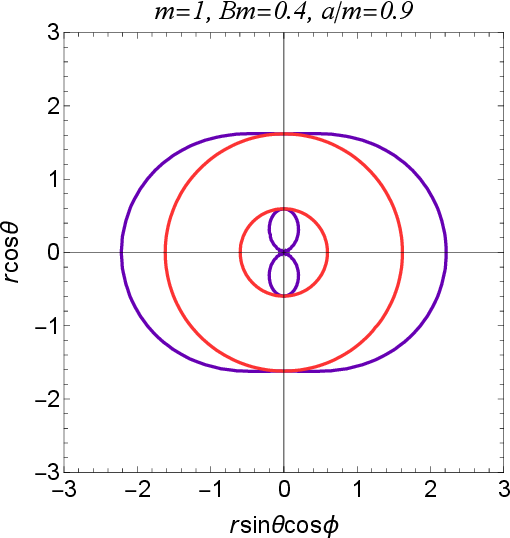}&
    \includegraphics[scale=0.5]{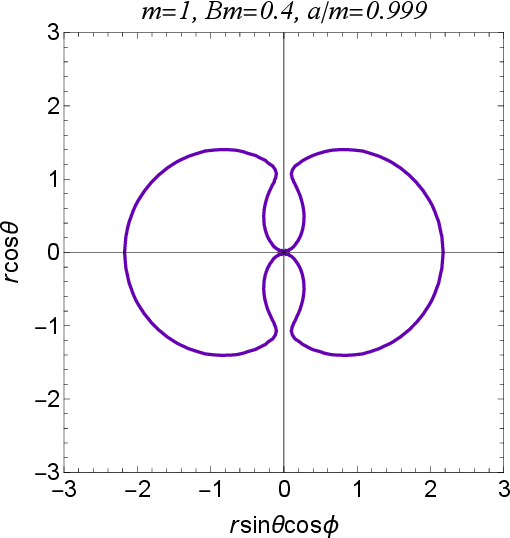}\\
    \includegraphics[scale=0.5]{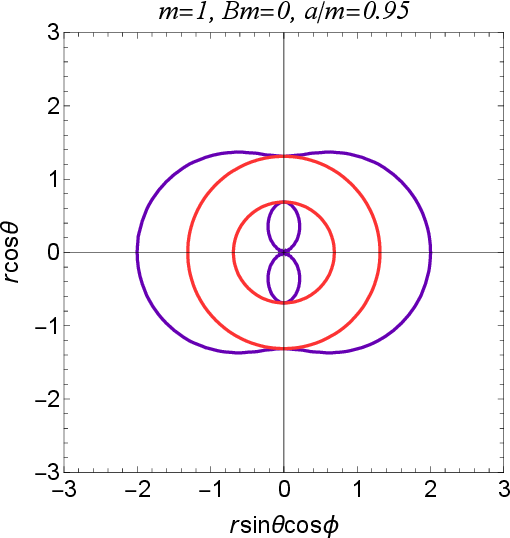}&
    \includegraphics[scale=0.5]{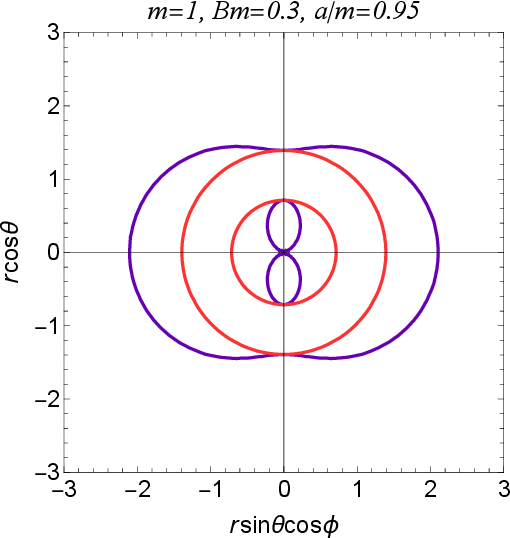}&
    \includegraphics[scale=0.5]{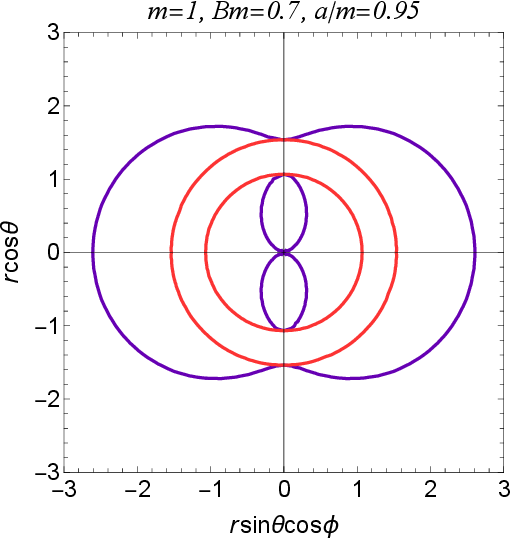}&
    \includegraphics[scale=0.5]{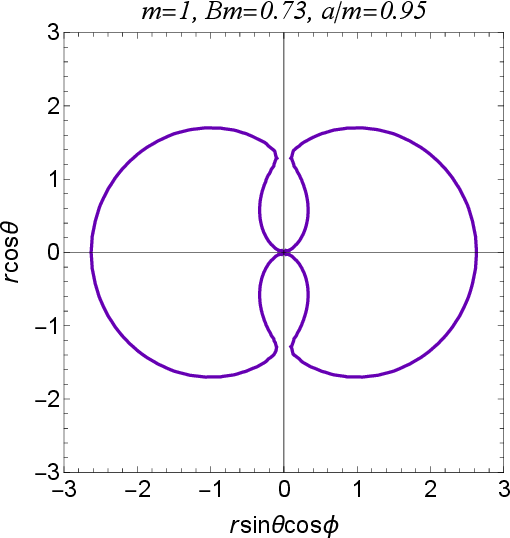}
    \end{tabular}
\caption{Cross-section of the event horizon ( red curve), stationary limit surface (SLS; blue curve), and ergoregion for KBRBHs with different magnetic field strengths $B$.  Horizons get disconnected as the spin parameter $a$ or $B$ increases. Horizontal deformation is caused by increased magnetic fields.  The ergoregion extends with both $a$ and $B$, while the SLS moves outward, indicating frame-dragging.}\label{ergo}
\end{figure*}

The magnetic deviation parameter $B$ presents several key modifications to the properties of Kerr black holes. An increase in $B$ restricts the maximum allowed spin $a$, making the attainment of extremality more difficult.
The ergoregion expands with increasing $B$ (cf.~Fig.\ref{ergo}), thereby modifying frame-dragging effects and the efficiency of energy extraction. Importantly, while for small $B$ the shadows remain nearly degenerate with those of Kerr, at larger $B$ the deviations become significant, though still within the present EHT observational uncertainties. Thus, $B$ acts as a genuine geometric deformation parameter encoding the backreaction of magnetic fields on black hole structure and observable signatures.

\section{Null Geodesics around Rotating Black Holes }\label{Sec3}
Assume the black hole is in front of an extended source of light. Depending on photon energy and angular momentum, source photons can scatter, capture, or move in unstable orbits \cite{Chandrasekhar:1985kt}. The marginally trapped photons that circle the black hole several times before reaching the distant observer form the black hole shadow barrier. Tracing back these photons' trajectory will give us the apparent shape and size of the black hole shadow, which, for the Kerr black hole, depends on the black hole parameters ($B, a$) and interestingly, the inclination angle ($\theta_o$) between the direction of a distant observer and the axis of rotation.

There is a significant deflection of light beams from background sources whose impact parameter is higher than the crucial value. They can reach an observer, while those with smaller impact parameters fall below the event horizon, generating a black region known as the \emph{shadow}, encircled by a bright photon ring. Synge (\citeyear{Synge:1966okc}) and Luminet (\citeyear{Luminet:1979nyg}) created a formula for the angular radius of the photon area surrounding a Schwarzschild black hole.  Bardeen (\citeyear{Bardeen:1973tla}) later investigated the shadow of a Kerr black hole and discovered that its spin affects the shape. The photon ring delineates the shadow and is determined by spacetime geometry, making its shape and size crucial for extracting black hole properties \citep{Afrin:2021imp, Kumar:2018ple} and disclosing near-horizon gravitational phenomena. Extensive analytical and numerical work has since explored shadows of various black hole types \citep{Vries2000TheAS,Shen:2005cw,Amarilla:2010zq,Yumoto:2012kz,Amarilla:2013sj,Atamurotov:2013sca,Abdujabbarov:2016hnw,Abdujabbarov:2015xqa,Cunha:2018acu,Mizuno:2018lxz,Mishra:2019trb,Shaikh:2019fpu,Kumar:2020yem,Afrin:2021imp,Afrin:2024khy,Islam:2022wck,KumarWalia:2022aop,Ghosh:2022kit,Afrin:2021wlj,Atamurotov:2024nre,Molla:2024lpt}, including for parameter estimation \citep{Afrin:2024khy,Vachher:2024fxs,Afrin:2021imp,Kumar:2018ple} and testing gravity theories \citep{Kramer:2004hd}.
We start with the Hamilton-Jacobi equation to analyze the null geodesics of photons in the KBRBH spacetime (Eq.~\ref{MKBH}) \citep{Carter:1968rr,Chandrasekhar:1985kt}: 
\begin{eqnarray} \label{HmaJam} \frac{\partial S}{\partial \lambda} = -\frac{1}{2}g^{\alpha\beta}\frac{\partial S}{\partial x^{\alpha}}\frac{\partial S}{\partial x^\beta}, \end{eqnarray} 
where $S$ is the Jacobi action and $\lambda$ is the affine parameter along the geodesics. 
Thus, the Jacobi action can be expressed as: \begin{eqnarray} S = -E t + L_Z \phi + S_r(r) + S_\theta(\theta), \end{eqnarray} where $S_r(r)$ and $S_\theta(\theta)$ are, respectively,  functions of the radial and angular coordinates. 

In general relativity, light (or photons) always follows \emph{null geodesics}, 
that is, paths for which the spacetime interval vanishes, \( ds^{2} = 0 \).
Under a conformal transformation of the metric, 
\[ g_{\mu\nu} \rightarrow \Omega^{-2}\tilde{g}_{\mu\nu}, \]
null geodesics remain invariant because the overall conformal factor 
\( \Omega^{-2} \) cancels out in \( ds^{2} = 0 \). 
Hence, the conformal factor does not alter the \emph{paths} of light rays, though it may affect affine parametrisation. Thereby, in the KBRBH spacetime, 
the conformal factor \( \Omega^{-2} \) does not influence photon motion. 
The photon dynamics depend only on the conformally related metric 
\( \tilde{g}_{\mu\nu} \), thereby we  neglect  \( \Omega^{-2} \) 
in analysing null geodesics, shadows, and related optical properties 
\cite{Wald:1984rg, Hawking:1973uf, Poisson:2009pwt,Bardeen:1973tla}.

The geodesic motion of photons around the black hole is necessary for the shadow formation. To study the same, we analyse the motion of a test particle in stationary and axially symmetric spacetime. Since metric (\ref{MKBH}) is independent of $t$ and $\phi$, these are the cyclic coordinates with corresponding killing vectors  $\chi_{(t)}^{\mu}=\delta _t^{\mu }$  and $\chi_{(\phi)}^{\mu}=\delta _{\phi }^{\mu }$, whose existence ensures conserved quantities for null geodesics: energy $E = -p_t$ and axial angular momentum $L_Z = p_{\phi}$, where $p_{\mu}$ is the photon's four-momentum. Additionally,in physically relevant Petrov type D spacetimes— such as the Kerr or Kerr–Newman families— the geometry admits a nontrivial second-rank Killing tensor that gives rise to Carter’s separability constant  $\mathcal{K}$, enabling separation of the Hamilton–Jacobi and related field equations \cite{Walker:1970un,Carter:1968rr}. While the Petrov type D classification is often associated with such separability properties, it does not by itself guarantee them; the existence of the Killing tensor (or equivalently, a Killing spinor) \citep{Carter:1968rr}.  The motion of the test particle, neglecting the back reaction, is governed by the rest mass $m_0$, total energy $E$, axial angular momentum $L_z$ and Carter constant $\mathcal{Q}$, which is related to the second-rank irreducible tensor field of hidden symmetry \citep{Carter:1968rr}. To obtain the geodesic equations in the first-order differential form, we utilise the Hamilton-Jacobi equation using the integral approach pioneered by Carter \citep{Carter:1968rr}, which for the metric (\ref{MKBH}) read \citep{Chandrasekhar:1985kt}

\begin{align}
\Sigma \frac{dt}{d\lambda}=&\frac{r^2+a^2}{Q}(E(r^2+a^2)-aL_{z})-\frac{a}{P}(aE\sin^2{\theta}-L_{z}),\label{32}\\
\Sigma \frac{d\phi}{d\lambda}=&\frac{a}{Q}(E(r^2+a^2)-aL_{z})-\frac{1}{P}(aE-\frac{L_z}{\sin^2{\theta}}),\label{33}\\
\Sigma \frac{dr}{d\lambda}=&\pm\sqrt{\mathcal{R}(r)}\ ,\label{req} \\
\Sigma \frac{d\theta}{d\lambda}=&\pm\sqrt{\Theta(\theta)}\ ,\label{theq}
\end{align}
where the effective potentials $\mathcal{R} (r)$ and ${\Theta}(\theta)$ for radial and polar motion are given by 
\begin{align}
\mathcal{R}(r)=&E^2\left[\Big((r^2+a^2)-a\xi \Big)^2-\Delta \Big({\eta}+(a-{ \xi})^2\Big)\Big(1+B^2r^2\Big)\right],\label{Rpot}\\
\Theta(\theta)=&E^2\left[\eta-\left(\frac{{ \xi}^2}{\sin^2\theta}-a^2 \right)\cos^2\theta + B^2\left(\frac{I_2}{I_1^2} - a^2\right)[({ \xi}-a)^2+\eta]\cos^2\theta\right]\ . \label{theta0}
\end{align}

The separability constant $\mathcal{K}$ is connected to the Carter constant $\mathcal{Q}$ by the equation $\mathcal{Q}=\mathcal{K}+(aE-L_z)^2$ \citep{Chandrasekhar:1985kt}. We introduce dimensionless quantities called impact parameters \citep{Chandrasekhar:1985kt}
\begin{eqnarray}
    \xi=\frac{L_z}{E} \;\; \text{,} \;\; \eta=\frac{\mathcal{K}}{E^2},
\end{eqnarray}
which are constant along the geodesics. In fact, there is a further relationship between $\eta$ and $\xi$ and the celestial coordinates. The allowable zone for photon mobility around a black hole is  $\mathcal{R}\geq 0$ and $\Theta\geq 0$ and the sign of $\dot{r}$ and $\dot{\theta}$ can be either positive or negative, chosen independently. At the turning points of motion, the sign change occurs, i.e., when $\mathcal{R}=0$ or $\Theta=0$ \citep{Chandrasekhar:1985kt}.

Our emphasis is on spherical photon orbits, which are spherical lightlike geodesics confined on a sphere with a constant coordinate radius $r$ characterized by $\dot{r}=0$ and $\ddot{r}=0$ \citep{Frolov:1418196,Chandrasekhar:1985kt}. It is crucial to examine the radial motion in order to determine the photons' circular orbits. The radial equation of motion (\ref{req}) can be rewritten as \cite{Wei:2013kza}
\begin{figure*}[hbt!]
\centering
\begin{tabular}{c c }
 
    \includegraphics[scale=0.7]{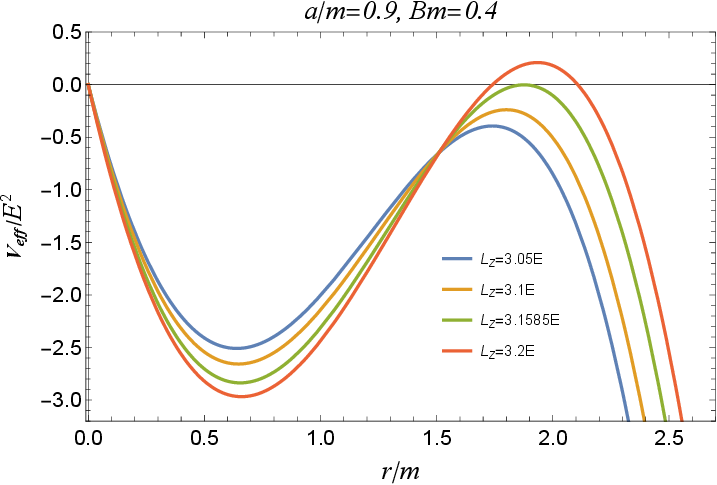}&
    \includegraphics[scale=0.7]{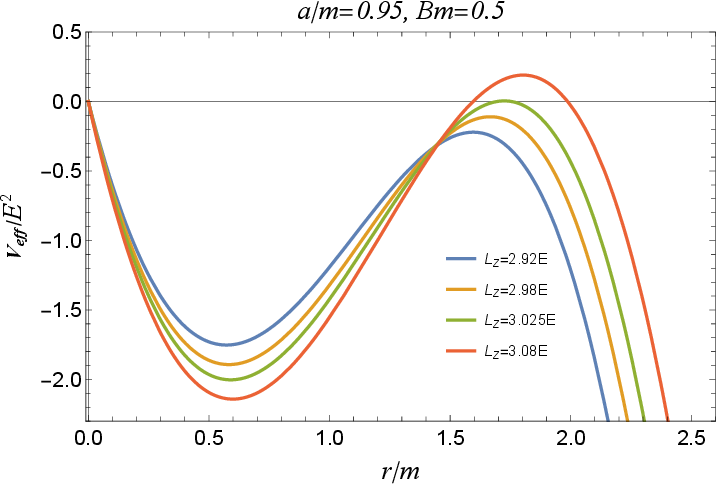}\\
    \end{tabular}

  \caption{Variation of the effective potential $V_{eff}$ for KBRBHs as a function of radial coordinate r with varying angular momentum $L_Z$ at $a/m=0.9$ (left) and $a/m=0.95$ (right) . }\label{veff}	
 \end{figure*}

\begin{equation}
\left(\Sigma \frac{d{r}}{d\lambda}\right)^2+V_{eff}(r)=0,
\end{equation}
where $V_{eff}$ represents the test particle's effective potential. We have shown the dependence of $V_{eff}$ on $L_Z$ in fig.~\ref{veff}. The unstable orbits experience a radial turning point at the maximum value of the effective potential ($\dot{r}= \ddot{r}=0$).

\begin{equation}
V_{eff}=\frac{\partial V_{eff}}{\partial r}=0 \;\; \;\; \mbox{or}\;\;  \; \mathcal{R}=\frac{\partial \mathcal{R}}{\partial r}=0.\label{vr} 
\end{equation}
While these orbits are only planar at the equatorial plane, generic unstable orbits, often referred to as spherical photon orbits, lie on a three-dimensional plane depending on the value of the Carter constant. Solving Eq.~(\ref{vr}) for impact parameters, yield

\begin{eqnarray}
\mathcal{R}=\mathcal{R}'=0 \,\, \text{and}\,\,\mathcal{R}''\leq 0.\label{unstableOrbit} 
\end{eqnarray}

Solving Eq.~(\ref{unstableOrbit})  yields the critical values of impact parameters ($\xi_{c}, \eta_{c}$) for the unstable orbits, which read as
\begin{align}
\xi_{c}=& \frac{(a^2+r^2)\Delta '(r)-4r \Delta (r) }{a \Delta '(r)} \nonumber\\
\eta_{c}=&\frac{r^2 \left(8 \Delta (r) \left(2 a^2+r \Delta '(r)\right)-r^2 \Delta '(r)^2-16 \Delta (r)^2\right)}{a^2 \Delta '(r)^2}\label{CriImpPara},
\end{align}
where $'$ denotes the derivative concerning the radial coordinate.

 The lensed "photon ring" is created when photons with a critical impact parameter equal to the critical photon sphere radius, $R_c$, are trapped in an unstable circular orbit. In the Kerr metric \citep{Kerr:1963ud}, $R_c$ changes depending on the photon’s orientation relative to the spin axis, leading to a non-circular cross-section \cite{Bardeen:1973tla}. Although this variation is less than 4\%, it is potentially detectable \citep{Takahashi:2004xh, Johannsen:2010ru}. The unstable photon orbits have been investigated extensively for black holes and naked singularities \citep{Wilkins:1972rs,Goldstein1974,Johnston:1974pn,Izmailov1979,Izmailov1980,Teo:2020sey}, which are the boundaries between light ray capture and its scattered cross-section. Two circular photon orbits can exist in the equatorial plane of axially symmetric spacetimes: those that move in the black hole's rotational direction and those that move oppositely- referred to as prograde and retrograde photons, respectively. The Lens-Thirring effect \citep{Bardeen:1975zz} causes the rotation of a black hole to drag the inertial frame to an observer at infinity, resulting in shortened orbits for prograde photons to account for excess angular momentum. Conversely, the retrograde ones would need to revolve at larger radii, since they had lost some angular momentum \citep{Bardeen:1972fi,Teo:2020sey}. Only when $\mathcal{\eta}_{c}>0$ do non-planar or three-dimensional photon orbits form; in contrast, when $\mathcal{\eta}_{c}=0$, photon orbits are planar and limited to the equatorial plane. 

The black hole appears optically as the black hole shadow surrounded by the bright photon ring
\citep{Synge:1966okc,Bardeen:1973tla,Luminet:1979nyg,Cunningham:1973tf}.  The shadow has made it easier to estimate and measure various black hole properties, including mass, spin angular momentum, and other hairs \citep{Kumar:2018ple}.  As a result, it can be used to test the no-hair theorem \citep{Cunha:2015yba} and general relativity in the strong-field regime \citep{Gott:2018ocn,Kumar:2020owy}. The photon region surrounding the black hole's event horizon is formed by combining all unstable spherical photon orbits, i.e. the separatrix between photon geodesics that fall into the event horizon and those that escape to spatial infinity.  We suppose that there are equally dispersed light sources at infinity, and photons arriving with all potential impact parameters are either caught by or scattered near the black hole. We additionally assumed a distant observer at an inclination angle of $\theta_0$ with the black hole's rotating axis. The perceived angular distances of the image measured from the line of sight in directions parallel and perpendicular to the projected axis of rotation black hole onto the celestial sphere, respectively, are the celestial coordinates ($X$,$Y$) of the shadow boundary at the observer's sky \citep{Hioki:2009na}.

\begin{figure*}[hbt!]
\centering
\begin{tabular}{l    r }
 
    \includegraphics[scale=0.45]{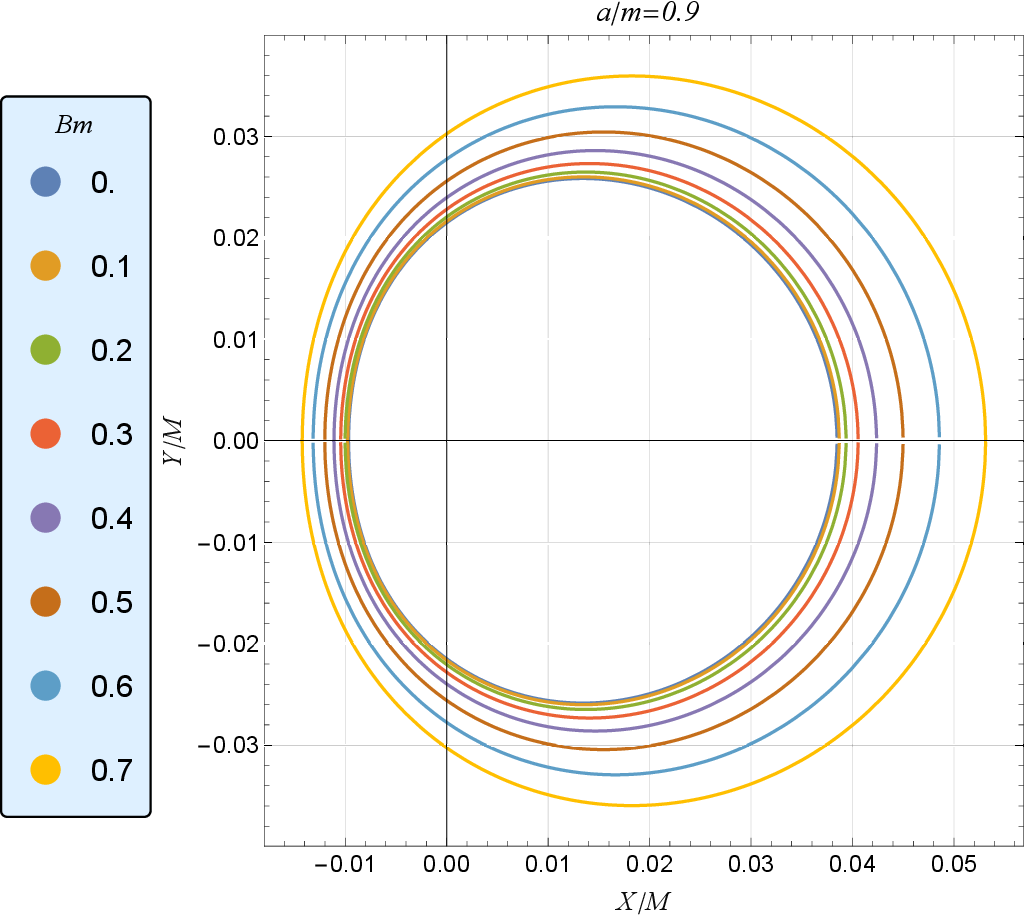}&
    \includegraphics[scale=0.44]{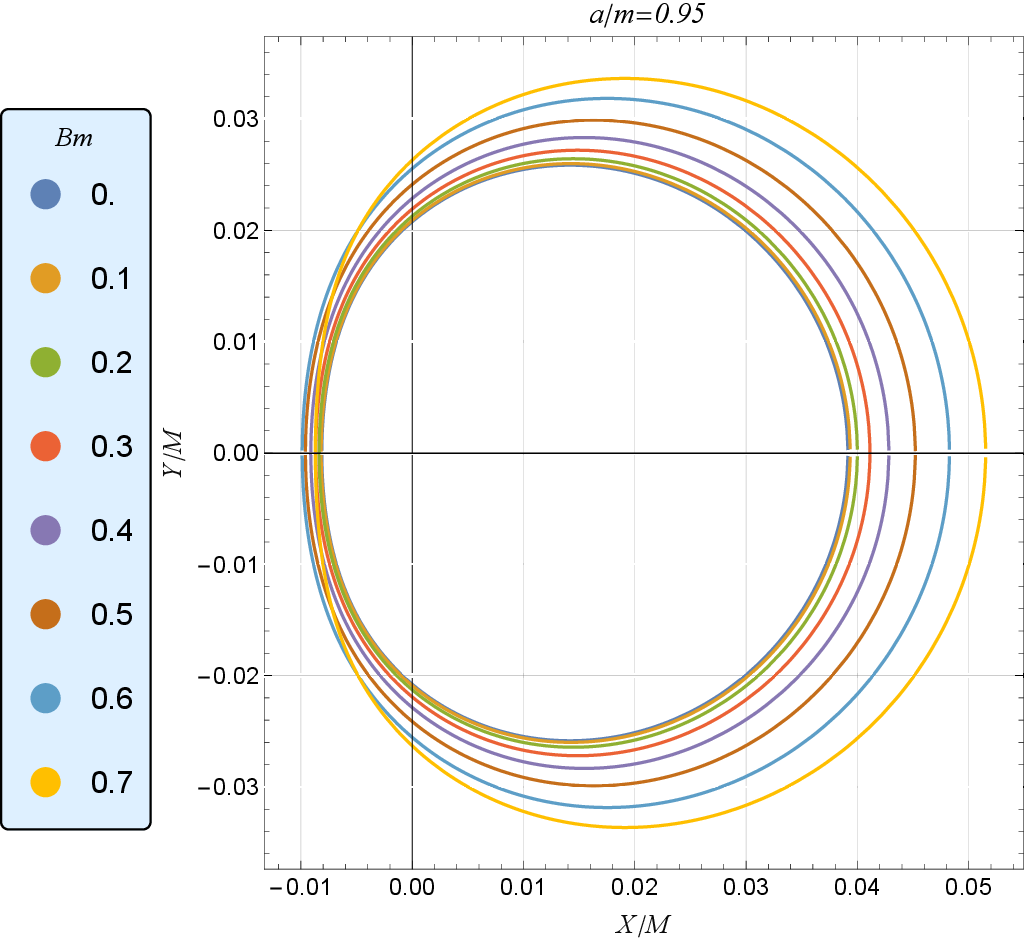}\\
    \end{tabular}

  \caption{Black hole shadow silhouettes of the KBRBH with various values of magnetic deviation parameter $B$ at $\theta_o=90^\circ$ , as shown for spin parameters $a/m = 0.9$ (left) and $a/m = 0.95$(right). As $B$ increases, the shadow size also increases and deviates from the Kerr case, revealing the effect of the external magnetic field on photon paths. The innermost contour represents the Kerr black hole. }\label{shadow}	
 \end{figure*}

 \begin{figure}[hbt!]
\includegraphics[scale=0.45]{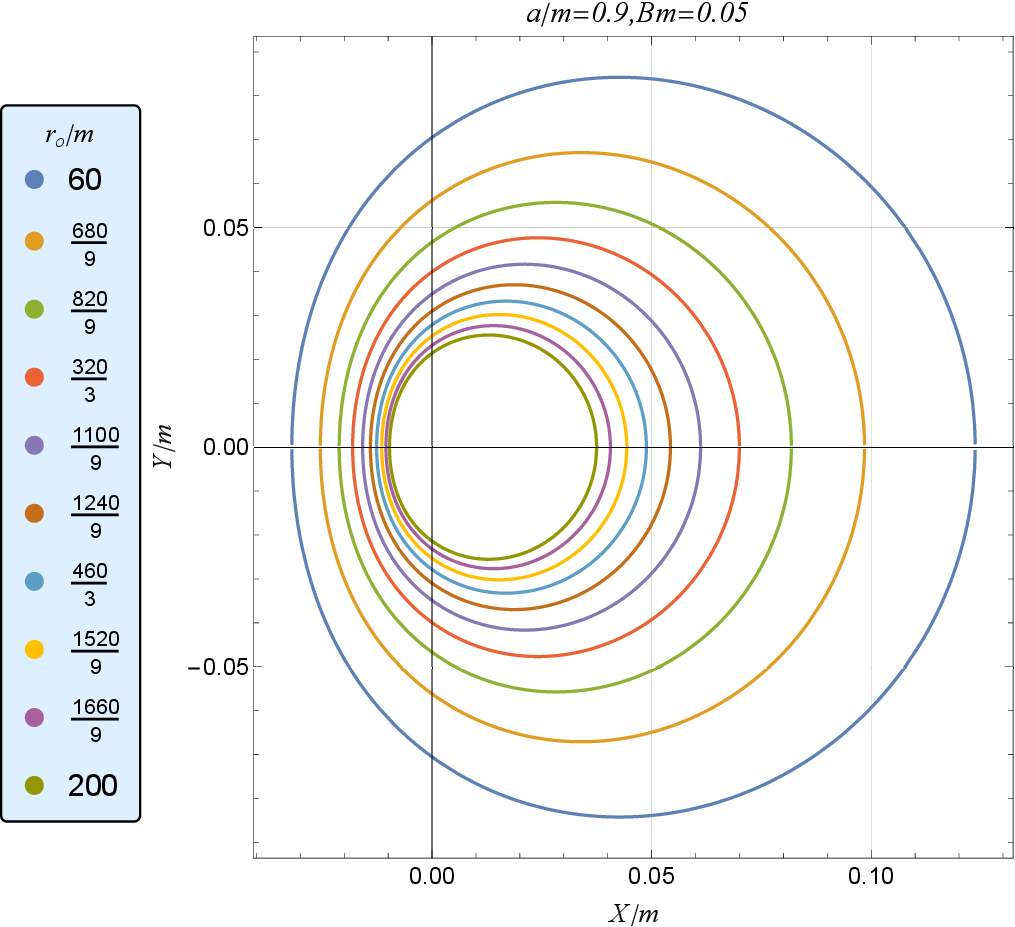}
\caption{\label{shadowMO} Shadows of KBRBHs as observed by a moving observer at different radial positions $r_O$, while the spin parameter $a$ and magnetic deviation parameter $B$ are kept fixed, at an inclination angle $\theta_o=90^\circ$}
\end{figure}
 \section{Shadow of  KBRBHs \label{Sec4}}
 Here, we explore in detail the black hole shadow of KBRBHs and analyze whether the presence of an external magnetic field introduces notable observational features. KBRBHs are exact rotating solutions to the Einstein--Maxwell equations with an external magnetic field parameter $ B $. These spacetimes are of Petrov type $D$ and are asymptotically non-flat due the magnetic field \cite{Podolsky:2025tle}. We know that a black hole shadow offers a direct way to probe the geometry of spacetime and help to estimate the physical parameters of black holes.

The shadow silhouettes surrounding a black hole are defined by the unstable circular photon orbits, which create a spharp contrast between the light and dark areas of the observer's sky. Under ideal conditions, the black hole properties and the underlying geometry dictate the size and shape of this shadow \cite{Falcke:1999pj, Chandrasekhar:1985kt}. However, these characteristics can be significantly influenced by including the magnetic fields or modified gravity effects. In a rotating geometry, assuming the direction of rotation to be counter-clockwise, the  prograde orbit photons (left side of the shadow) are closer to the event horizon as compared to the retrograde photons (right side)  \cite{Hioki:2009na}. This assyemetry arises due to the fact that prograde and retrograde orbit photons do not experience the same effective potential \cite{Bambi:2008jg}. With an increase in the rotation parameter $a$, the prograde photon orbit shrinks rapidly as compared to the retrograde one resulting in an assymetrical and distorted shadow. At high spins ($a \gtrsim 0.5$), the shadows shift from the circular symmetry of a non-rotating black hole due to a reduction in the perpendicular direction to the spin axis \cite{bardeen1973, Chandrasekhar:1985kt}.

 Additional distortions are introduced to KBRBHs by the presence of the magnetic field parameter $B$ resulting in the modification of shadow shape, shift in the centroid and compression of shadow boundary. We cannot place the observer at spatial infinity as the spacetime is not asymptotically flat, hence a static observer is considered at a large finite radial distance $\mathbf {r_O=200m}$ in a way that $r_O \gg r_+$. The observer is considered to be in the equatorial plane ($\theta_O = \pi/2$) so that the spin-induced effects are maximum.

We exploit the standard technique of ray tracing from the observer backwards in time to study photon trajectories. These photons either fall into the black hole or go to the background source. The boundary between these two scenarios define the shadow. An orthonormal tetrad is created at the observer's location, according to Grenzebach \cite{Grenzebach:2014fha}, as follows:

\begin{eqnarray}\label{35}
e_0=\frac{(r^2+a^2)\partial_t+a\partial_{\phi}}{\sqrt{\Sigma\Delta}},\;\;\;\;e_1=\frac{1}{\sqrt{\Sigma}}\partial_{\theta},\\
e_2=-\frac{\partial_{\phi}+a\sin^2{\theta}\partial_t}{\sqrt{\Sigma}\sin\theta},\;\;\;\; e_3=-\sqrt{\frac{\Delta}{\Sigma}}\partial_r.
\end{eqnarray}

A tangent vector to a light ray $\lambda(\tau)$, with affine parameter $\tau$, reads
\begin{equation}
\dot{\lambda}(\tau)=\dot{t}\partial_t+\dot{r}\partial_r+\dot{\theta}\partial_{\theta}+\dot{\phi}\partial_{\phi}, \label{lambda1}
\end{equation}
and it can be expressed at the observer's position as
\begin{equation}
\dot{\lambda}(\tau)=\alpha(-e_0+\sin\Phi\cos\Psi e_1 +\sin\Phi\sin\Psi e_2 +\cos\Phi e_3 ).\label{lambda2}
\end{equation}
Using conserved quantities $E$ and $L_Z$, the scalar factor $\alpha$ is calculated as
\begin{equation} \alpha=\frac{aL_Z-(r^2+a^2)E}{\sqrt{\Sigma\Delta}}.\label{alpha1} 
\end{equation}

Corresponding to the photon's direction in the observer's sky, the celestial coordinates $\Phi$ and $ \Psi $ are given by \cite{Grenzebach:2014fha}
\begin{eqnarray}
\sin\Psi(r_O,r_p) &=&\left.\left( \frac{\xi-a}{\sqrt{(a-\xi)^2+\eta}}\right)\right|_{r=r_O},\\ \label{Psieq}
\sin\Phi(r_O,r_p) &=&\left.\left(\frac{\sqrt{\Delta[(a-\xi)^2+\eta]}}{(r^2+a^2-a\xi)}\right)\right|_{r=r_O},\label{Phieq}
\end{eqnarray}
where $ \xi $ and $\eta$ are the conserved quantities corresponding to the impact parameters. We introduce Cartesian coordinates $ X$ and $ Y$ in the observer’s sky, to  visualize the shadow as:
\begin{eqnarray}
X(r_O,r_p)=-2\tan\left(\frac{\Phi(r_O,r_p)}{2}\right)\sin(\Psi(r_O,r_p)), \label{X11}\\
Y(r_O,r_p)=-2\tan\left(\frac{\Phi(r_O,r_p)}{2}\right)\cos(\Psi(r_O,r_p)).\label{Y1}
\end{eqnarray}

Plotting $X$ versus $Y$ gives the shadow of the black hole as seen by the observer at $r_O$. The shadow shifts in the direction of black hole rotation, hence indicative of increased asymmetry caused by the magnetic field. Unlike Kerr-plus-plasma models, which only impact radiation transfer, KBRBHs add magnetic backreaction into spacetime geometry via the parameter $B$. Measuring shadow variations caused by $B$ would offer direct constraints on horizon-scale magnetic fields in SMBHs, complementing those predicted from polarization and jet-power  \cite{EventHorizonTelescope:2019dse,EventHorizonTelescope:2022wkp}. The next-generation Event Horizon Telescope (ngEHT) and future space-VLBI missions, aiming for \%-level precision in shadow diameter and asymmetry \cite{Ayzenberg:2023hfw}, may identify such effects, which are still beyond present EHT resolution.

Because the KBRBH spacetime is not asymptotically flat, we place the observer at a large but finite radius $r_O$ and construct an orthonormal tetrad at the observation event to map photon momenta to celestial coordinates $(X,Y)$ \cite{Grenzebach:2014fha}. As the observer moves inward (decreasing $r_O$), finite–distance lensing and relativistic effects amplify the apparent angular size of the shadow and enhance its spin-induced asymmetry. In practice, the boundary compression on the approaching (prograde) side and the centroid shift in the direction of rotation both increase as $r_O$ decreases, while $a/m$ and $Bm$ are held fixed \cite{Hioki:2009na}. In the magnetized case, $B$ enlarges the shadow and strengthens these  (cf.~Fig.\ref{shadow}), so the trends with $r_O$ add to the intrinsic $B$–driven distortion and centroid shift already present at fixed spin \cite{Podolsky:2025tle}. These outcomes are calculated by ray-tracing null geodesics backward from the moving observer’s sky using the tetrad-based angles $(\Phi,\Psi)$ and their projection to $(X,Y)$, Eqs.~\ref{35}--\ref{Y1}, applied at finite $r_O$ \cite{Grenzebach:2014fha}. Overall, Fig.~\ref{shadowMO} illustrates that decreasing $r_O$ acts similarly to increasing $B$: both make the shadow larger and more asymmetrical, with a stronger displacement of the silhouette. The behaviour of the shadow plots in Fig.~\ref{shadowMO} is also consistent with results reported in Refs. \cite{Afrin:2021ggx,Ahmed:2025boj}.

 \section{Parameter Estimation Using Shadow Observables: Kumar-Ghosh Method}\label{Sec5}
 The SMBHs M87* and Sgr A*'s observed images match the  Kerr black hole mentioned in general relativity. Nevertheless, black holes in modified theories of gravity were not mentioned in the EHT observation \citep{EventHorizonTelescope:2019dse,EventHorizonTelescope:2019ths,EventHorizonTelescope:2019ggy}, which is what we plan to investigate using the calculated parameters of our KBRBHs. We assume the observer is in the equatorial plane  ($\theta_0=\pi/2$) for parameter estimation. A black hole's shadow's shape and size can furnish necessary information about the black hole's properties and the underlying geometry of spacetime. Hioki and Maeda \citep{Hioki:2009na} suggested two shadow observables: the shadow radius $R_s$, which defines the approximate size, and the distortion parameter $\delta_s$, which quantifies the shadow's deviation from circularity. These observables allowed numerical estimations of black hole parameters, particularly the inclination angle and spin value $a$.  This method was then expanded to include analytical estimations, increasing its application (Cf.~\citet{Tsupko:2017rdo}).

In addition to that, it was highlighted by  \citet{Tsukamoto:2014tja} that variations caused by general relativity could be reflected in the shadow asymmetries, hence suggesting a method by which shadows of Kerr black holes can be distinguished from those resulting from modified theories of gravity.
These methods, however, rely primarily on assumptions like certain symmetries in the shape of shadow (near-circularity or equatorial reflection symmetry) when in reality the shadows may likely become irregular and asymmetric, specifically in the presence of deviations due to modified gravity, surrounding matter, or magnetic fields  \citep{Schee:2008kz, Johannsen:2015qca, Abdujabbarov:2015xqa, Younsi:2016azx}.

Also, significant noise, instrumental resolution limits, and image reconstruction artefacts are present in the observational data from the EHT, which can cause the shadow to distort considerably from the idealised symmetric form, making it more challenging to use standard circularity-based observables accurately \citep{Abdujabbarov:2015xqa, Kumar:2018ple, EventHorizonTelescope:2019dse, EventHorizonTelescope:2022wkp}. To overcome this, a more reliable collection of observables was proposed by \citet{Kumar:2018ple}. These observables can accurately describe non-circular and asymmetric shadow boundaries containing 
the total shadow area $A$ and oblateness $D$, which are, respectively, defined as follows:

\begin{eqnarray}
A &=& 2\int{Y(r_p) \, dX(r_p)} = 2\int_{r_p^{-}}^{r_p^+} \left( Y(r_p) \frac{dX(r_p)}{dr_p} \right) dr_p, \label{eq:Area}
\end{eqnarray}
\begin{eqnarray}
D &=& \frac{X_r - X_l}{Y_t - Y_b}, \label{eq:Oblateness}
\end{eqnarray}
where $(X_l, X_r)$ and $(Y_b, Y_t)$ represent the extremal coordinates of the shadow in the image plane—specifically, the leftmost, rightmost, bottommost, and topmost points of the shadow boundary.

 \begin{figure*}[hbt!]
		\begin{tabular}{c c}
			\includegraphics[scale=0.7]{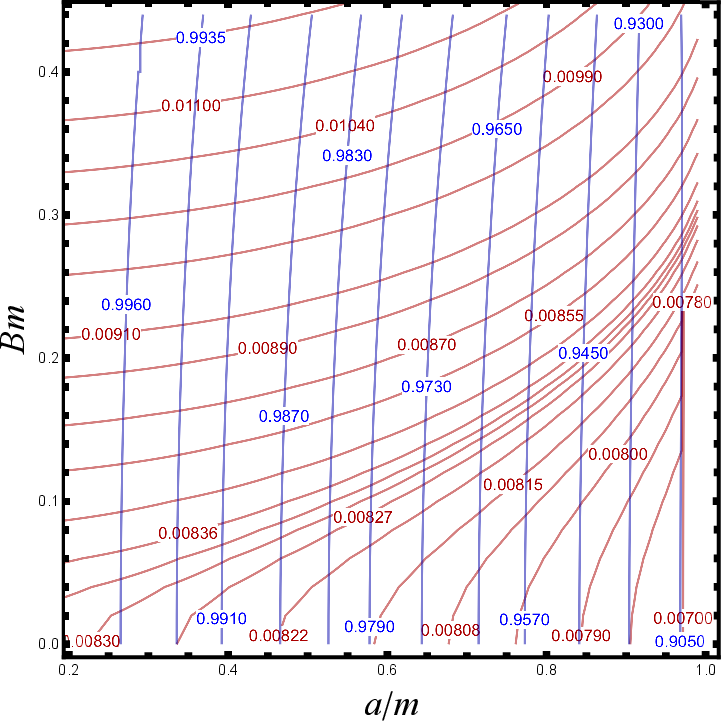}&
            \includegraphics[scale=0.7]{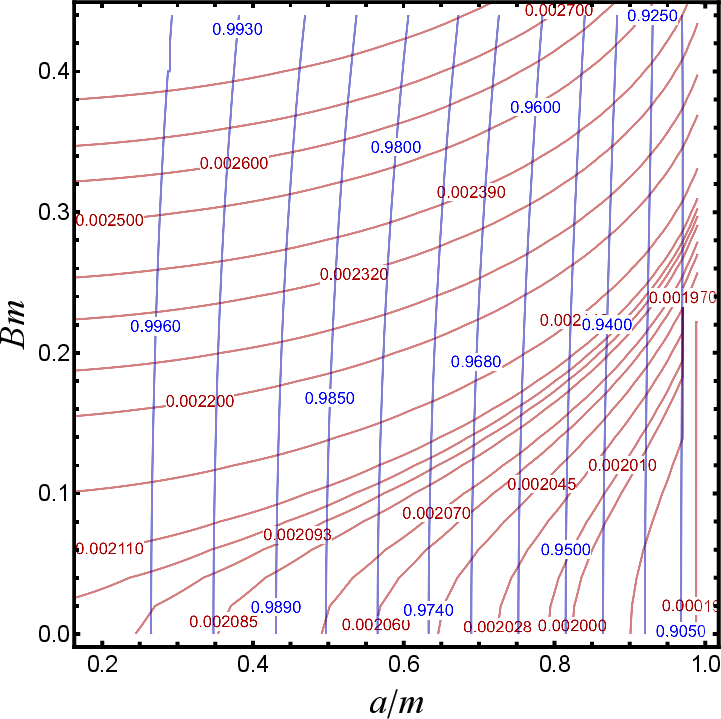}\\
		\end{tabular}
	\caption{\textbf{Parameter estimation using shadow observables. Contour plots in the KBRBH parameter space $(a, B)$ at $r_O = 100m$ (left) and $r_O = 200m$ (right) for an inclination angle $\theta_O = 90^\circ$. \textcolor{red!50!black}{Red contours} represent curves of constant shadow area $A$, while \textcolor{blue}{blue contours} correspond to curves of constant oblateness $D$. The intersection points of the $A$ and $D$ contours indicate the $(a, B)$ parameter values that best reproduce the observationally inferred shadow characteristics.}}
	\label{parameterestimation2}
\end{figure*}

For an equatorial observer ($\theta=\pi/2$), the oblateness $D$ is a natural measure of deviation from circularity. It ranges from $D=1$ for a perfectly circular shadow, such as in the non-rotating Schwarzschild black hole ($a=0$), down to $D = \sqrt{3}/2$ for the extremal Kerr case ($a=m$) \citep{Tsupko:2017rdo}. These observables are specifically helpful for studying KBRBHs, where a magnetic field parameter $B$ introduces further deviations from axial symmetry.

Here, we estimate the spin parameter $a$ and the magnetic field strength $B$ of the KBRBH spacetime using the shadow observables $A$ and $D$. We get a strong tool for parameter estimation by creating contour maps of constant $A$ and $D$ in the ($a$,$B$) parameter space. Precise estimates of the black hole parameters are obtained from the intersection sites of the contours corresponding to observationally inferred values of $A$ and $D$. The intersection points of the contours corresponding to observationally inferred values of $A$ and $D$ yield precise estimates of the black hole parameters.
\textbf{Figure~\ref{parameterestimation2} has been drawn using the data from Fig.~\ref{shadow}, and this ensures full consistency with Fig.~\ref{shadow}. The red contours denote the variation of the shadow area, whereas the blue contours correspond to the distortion parameter in the ($a$,$B$) plane. These contours provide a clear visualisation of how the black hole shadow’s size and shape evolve with the magnetic field $B$.}
 The intersection points show which parameter values match the specified shadow observables. In particular, this method permits limits on $a$ and $B$ even when the black hole shadows are asymmetric.  The obtained parameter values are shown in Tables~\ref{parameter_table1} and \ref{parameter_table2}, demonstrating the efficiency of this approach in real-world observational scenarios like those of EHT probes.
It is important to note that the KBRBH spacetime, although useful in exploring the effects of an external magnetic field on photon motion, does not share the same asymptotic structure as the Kerr black hole. In particular, the KBRBH  remains asymptotically uniform in the magnetic field. In contrast, in realistic astrophysical systems, the magnetic field strength is anticipated to decay at large distances from the black hole. Hence, the KBRBH  spacetime cannot accurately describe the magnetized environment of a black hole on very large scales. Consequently, our parameter estimation method based on the shadow analysis is physically significant only for observers relatively close to the black hole, typically within distances of order $\sim 100r_{+}$. 

 The magnetic parameter $B$ introduces degeneracies with the black hole spin $a$ and the observer's inclination angle $\theta_{0}$. Both $a$ and $B$ can grow and distort the shadow in comparable ways, while the inclination angle also determines the degree of asymmetry \cite{Bardeen:1973tla,Luminet:1979nyg,Hioki:2009na}. As a result, different combinations of $(a, B)$ might produce almost identical shadow forms, complicating parameter extraction from observations. Current EHT images of M87* and Sgr A* are compatible with Kerr-like shadows within $\sim10\%$ uncertainties \cite{EventHorizonTelescope:2019dse,EventHorizonTelescope:2022wkp}, indicating that deviations owing to small $B$ stay hidden within current error bands. However, for bigger $B$, the associated distortions become considerable and may be controlled by the future ngEHT target to reach \%-level precision in shadow measurements \cite{Ayzenberg:2023hfw}.

\begin{table*}
 
\begin{tabular}{|c c c c| }
\hline
A & D & $a/m$ & $Bm$ \\
\hline\hline 
0.0012 & 0.996 & 0.2889 & 0.4206 \\
\hline 
0.0012 & 0.991 & 0.427 & 0.4321 \\
\hline 
0.0104 & 0.996 & 0.279 &0.3344 \\
\hline 
0.0104 & 0.987 & 0.4905 & 0.3556 \\
\hline 
0.0104 & 0.965 & 0.7443 & 0.4041\\
\hline
0.0095 & 0.9935 & 0.3464 &  0.2669 \\ 
\hline 
0.0095 & 0.979 &0.5954& 0.2989 \\
\hline 
0.0095 & 0.945 & 0.8582 &  0.3697 \\
\hline 
0.0087& 0.996 & 0.2676&  0.157 \\
\hline 
0.0087 & 0.987 & 0.4727 & 0.1784 \\
\hline 
0.0087 & 0.955 & 0.7231& 0.2274 \\
\hline 
0.00843 & 0.9935 & 0.338  & 0.1007\\
\hline
0.00843& 0.983 & 0.5295  & 0.1304\\ 
\hline 
0.00843 & 0.965 & 0.7197  & 0.1752 \\
\hline 
0.0083 & 0.996 & 0.263  & 0.02349\\
\hline 
0.0083 & 0.987 & 0.4657 & 0.07291 \\ 
\hline 
0.0083 & 0.979 & 0.5802 & 0.1028 \\ 
\hline

\end{tabular}
\caption{Estimated values of the KBRBH parameters $B$ and $a$ for $r_O = 100m$, derived from the shadow observables—the area $A$ and oblateness $D$—at an inclination angle $\theta = 90^\circ$ (cf. Fig.~\ref{parameterestimation2} (left)). The estimates correspond to the intersection points of constant-$A$ and constant-$D$ contours in the $(a, B)$ parameter space}\label{parameter_table1}
\end{table*}  
 
\begin{table*}
 
\begin{tabular}{|c c c c|}
\hline
A & D & $a/m$ & $Bm$ \\
\hline\hline 
0.00285 & 0.996 & 0.2836 & 0.387 \\
\hline 
0.00285 & 0.985 & 0.5317 & 0.4098 \\
\hline 
0.0026 & 0.996 & 0.2786 &0.3284 \\
\hline 
0.0026 & 0.989 & 0.4535 & 0.343 \\
\hline 
0.0026 & 0.968 & 0.7189 & 0.3879\\
\hline
0.00232 & 0.993 & 0.3575 &  0.2348 \\ 
\hline 
0.00232 & 0.974 &0.6481 & 0.2747 \\
\hline 
0.00232 & 0.94 & 0.8783 &  0.347 \\
\hline 
0.0022& 0.996 & 0.2673 &  0.1607 \\
\hline 
0.0022 & 0.98 & 0.5732 & 0.1979 \\
\hline 
0.0022 & 0.925 & 0.9263 & 0.3121 \\
\hline 
0.00214 & 0.989 & 0.4347  & 0.1278\\
\hline
0.00214& 0.974 & 0.6378  & 0.1668 \\ 
\hline 
0.00214 & 0.925 & 0.9263  & 0.273 \\
\hline 
0.0021 & 0.996 & 0.2647  & 0.06613\\
\hline 
0.0021 & 0.98 & 0.5673 & 0.1059 \\ 
\hline 
0.0021 & 0.96 & 0.7577 & 0.163 \\ 
\hline

\end{tabular}
\caption{Estimated values of the KBRBH parameters $B$ and $a$ for $r_O = 200m$, derived from the shadow observables—the area $A$ and oblateness $D$—at an inclination angle $\theta = 90^\circ$ (cf. Fig.~\ref{parameterestimation2} (right)). The estimates correspond to the intersection points of constant-$A$ and constant-$D$ contours in the $(a, B)$ parameter space}\label{parameter_table2}
\end{table*}

\section{Energy emission rate}
\label{EE rate}
We then examine the energy emission rate for the rotating KBRBH, which is an important metric that links possible astrophysical signatures to black hole thermodynamics. In the high-frequency regime ($\omega m \gg 1$), the absorption cross-section for massless fields approaches a constant limiting value \cite{Sanchez:1976fcl,Unruh:1976fm}. For a distant observer, this limit is fundamentally governed by the unstable photon orbits constituting the photon sphere. While the exact limiting value exhibits dependence on spin and field characteristics for rotating black holes, it remains well-approximated by the geometrical cross-section of the photon sphere, directly related to the shadow radius $R_s$ \cite{Decanini:2010fz,Perlick:2021aok}.
 In the high-frequency limit ($\omega m \gg 1$), the absorption cross section of a black hole is well known to asymptotically approach a constant value \cite{Sanchez:1976fcl,Unruh:1976fm}, given by

\begin{equation}
\sigma_{lim} \approx \pi R_{s}^2, 
\end{equation}
where $R_{s}$ is the KBRBH shadow radius, which plays an essential role in determining the high-energy absorption cross section perceived by a distant observer \cite{Decanini:2010fz,Perlick:2021aok}. This limiting value, obtained in terms of photon geodesics, coincides with the geometrical cross section of the photon sphere, and its interpretation has also been confirmed from wave scattering analyses \cite{Mashhoon:1973zz,Dasgupta:1996qu}. 

The black hole shadow is responsible for the high-energy absorption cross-section for a distant observer. The aforementioned limiting constant value can be analysed for wave theories and is calculated in terms of geodesics. The limiting constant value for a black hole with a photon sphere is equal to the photon sphere's geometrical cross-section. 

On using this limiting cross section, the rate of energy emission can be expressed in terms of the Hawking spectrum \cite{Page:1976df,Wei:2013kza} as
\begin{equation}
\frac{d^2 E(\omega)}{d\omega\, dt} = \frac{2 \pi^2 R_{s}^2}{e^{\omega/T_H} - 1} \, \omega^3 ,
\end{equation}

\begin{figure*}
 \includegraphics[width=0.48\linewidth]{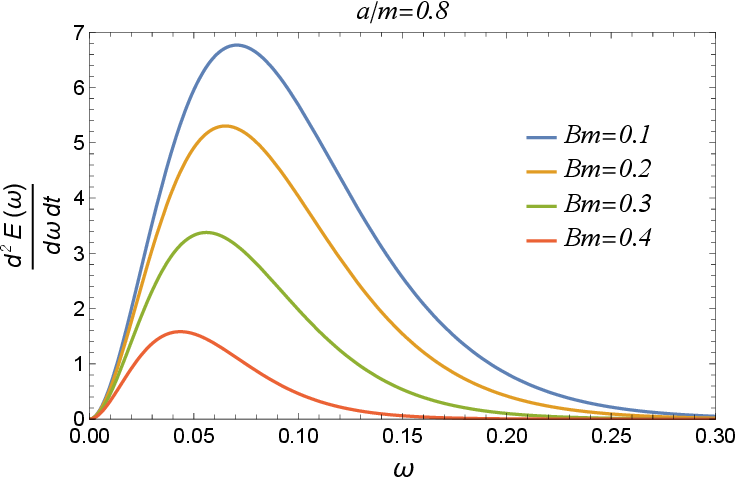}
 \includegraphics[width=0.48\linewidth]{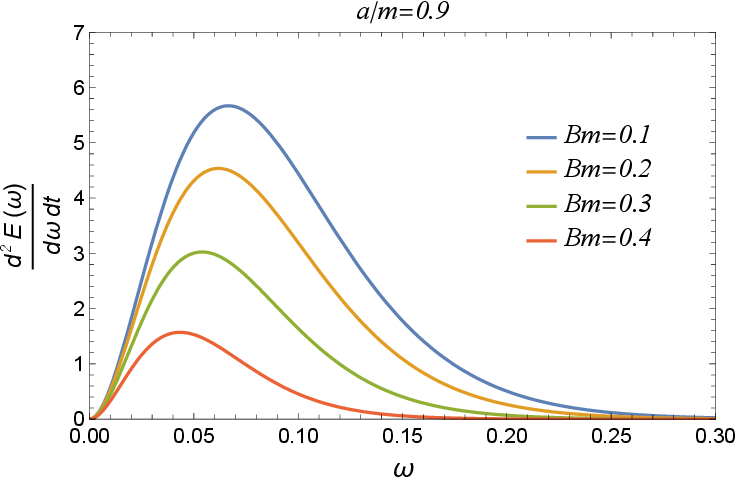}
\caption{\label{ee} Plots showing the behavior of the energy emission rate versus frequency, $\omega$ for $a/m=0.8$ (Left) and $a/m=0.9$ (Right).}
\end{figure*}

In the Figure~\ref{ee},  we have plotted the energy emission rate $\left(d^2 E(\omega)/d\omega\, dt\right)$ against frequency $\omega$ for  $a/m=0.8$ and $a/m=0.9$ with varying the magnetic deviation parameter $B$. For small $B$, the emission rate exhibits a higher peak, whereas increasing $B$ suppresses the maximum emission. Such a behaviour is a consequence of the decrease in $T_H$ due to magnetic field $B$, echoing the fact that stronger magnetic fields lower the effective thermal radiation output.

It turns out that other modified theories black hole spacetimes, including Einstein–Gauss–Bonnet gravity \cite{Konoplya:2020bxa}, quantum-corrected black holes \cite{Kumar:2020ltt}, and regular black holes \cite{Toshmatov:2015wga}, have also shown similar suppression effects. In these cases, extra parameters change the near-horizon geometry, which lowers the temperature and emission rate. In this way, the parameter $B$ encodes additional physical effects in the energy emission spectrum in a manner similar to these deformations. 
Our results on the ernergy-emisions rate show that the KBRBH spacetimes influence the black hole's Hawking radiation spectrum in addition to changing the size and shape of the shadows. This suggests a potential path for future high-precision observations of black hole shadows to probe horizon-scale magnetic fields \cite{Ayzenberg:2023hfw}.

\section{Conclusions}\label{Sec6}
There is an urgent need to test deviations from the Kerr geometry using high-resolution EHT observations of black hole shadows. While the Kerr solution remains the foundation of astrophysical black hole modelling in general relativity, it ignores true astrophysical factors like magnetic fields that commonly occur in accretion environments. The KBRBH, with additional magnetic deviation parameter $B$, provides a theoretically consistent and observationally suitable extension of the Kerr solution. Given that surrounding matter can introduce subtle but measurable modifications to black hole shadows, we aimed to explore whether KBRBHs can serve as viable candidates for the supermassive black holes. By estimating the KBRBH parameters using shadow observables, this work contributes to the broader goal of probing gravity in the strong-field regime.

Motivated by this, we have conducted a comprehensive investigation into the shadows of KBRBHs, characterized by a deviation parameter $B$.  We have theoretically investigated the shadows and observables of KBRBHs, which are exact solutions to the Einstein-Maxwell equations incorporating an external, uniform magnetic field parameterised by $B$. Our primary goal was to develop a framework for measuring the effects of an active backreaction magnetic field on the spacetime geometry on astrophysical observables and shadow.

By analyzing null geodesics and computing shadow observables, we demonstrated how an external magnetic field influences the horizon structure and the shadow morphology of rotating black holes.  We estimate the parameters of KBRBHs, which are rotating solutions to the Einstein--Maxwell equations with an external magnetic field parameter $ B $. The spacetimes are of Petrov type $ D $ and are asymptotically non-flat due to the presence of the magnetic field. In contrast to the standard Kerr solution, this external magnetic field alters the geometry.   We investigate how the KBRBH spacetime influences photon orbits and shadows.

We summarise our main findings as follows:

\begin{itemize} 

\item The size and shape of a black hole's shadow are altered by the addition of a magnetic field, $ B$. We found that the shadow gets enlarged and more deformed in shape as the magnetic field strength increases. The analysis and all plots shows  good agreement with the results reported in Refs. \cite{Wang:2025vsx,Zeng:2025tji}, and similar behavior of shadows has also been reported in related studies \cite{Afrin:2021ggx,Ahmed:2025boj}, with respect to the moving observer.

 \item Using the separability of the Hamilton--Jacobi equation in the KBRBH spacetime, we derived null geodesics and critical impact parameters for unstable photon orbits with aid of effective potential. These were used to compute the shadow contour for various values of spin $ a $ and magnetic field strength $ B $.
    
\item A key outcome of this study is the development of a ready-to-use methodology for constraining $B$ should future data warrant it. By employing the Kumar-Ghosh formalism---characterizing the shadow via its area $A$ and oblateness $D$---we have created a degenerate-breaking tool. The contour plots of these observables in the $(a, B)$ parameter space provide a direct map from a measured shadow morphology to estimates of both the spin $a$ and the magnetic field strength $B$.

\item The energy emission rate study shows that the Hawking radiation is greatly altered by the magnetic field parameter $B$. We find that the black hole's shadow size is the main determinant of its absorption cross-section in the high-frequency regime. Interestingly, when the magnetic field intensity $B$ rises, the energy emission rate is suppressed. The backreaction is the cause of this suppression since it lowers the black hole's Hawking temperature and, as a result, its effective thermal radiation output.

\end{itemize}
In summary, this study serves as a crucial preparatory step in the search for astrophysical magnetic fields beyond the test-field approximation. We have quantified the theoretical imprints of a backreacting magnetic field on black hole shadows and developed a robust pipeline for parameter estimation. 

 Lastly, the KBRBH  is a powerful tool for exploring black hole physics that goes beyond what Einstein's general relativity explains. To understand gravity in its most extreme form and test how quantum mechanics works near a black hole, we need to keep studying these phenomena with both new observations and continued theoretical work. To evaluate KBRBH scenarios and establish quantitative limitations on near-horizon magnetic fields in SMBHs such as M87* and SgrA*, the ngEHT provides an efficient approach.
 
 Our work thus provides the necessary theoretical toolkit and quantitative predictions that will be essential for interpreting future high-precision observations and ultimately determining the role of strong magnetic fields in shaping the spacetimes of astrophysical black holes.

\section*{Acknowledgements}
S.G.G. gratefully acknowledges the support from ANRF through project No. CRG/2021/005771. 

%%%%%%%%%%%%%%%%%%%%%%%%%%%%%%%%%%%%%%%%%%%%%%%%%%
\section*{Data Availability}

We have not generated any original data in the due course of this study, nor has any third-party data been analysed in this article.

\bibliographystyle{apsrev4-1}
\bibliography{MKBH}
\end{document}